\documentstyle[aps,prbbib,psfig]{revtex}

\newcommand{\bvec}[1]{{\mathbf #1}}

\begin{document}
\draft
\widetext
\title{Aging dynamics of quantum spin glasses of rotors}
\author{Malcolm P. Kennett$^{a}$, Claudio Chamon$^{b}$, and Jinwu Ye$^{c}$}

\address{ 
$^a$ Department of Physics, Princeton University, Princeton, NJ 08544
\\
$^b$ Department of Physics, Boston University, Boston, MA
02215 \\
$^c$ Department of Physics, University Houston, TX 77204}

\date{\today}

\twocolumn[\hsize\textwidth\columnwidth\hsize\csname@twocolumnfalse\endcsname

\maketitle

\begin{abstract}
We study the long time dynamics of quantum spin glasses of rotors
using the non-equilibrium Schwinger-Keldysh formalism. These models are
known to have a quantum phase transition from a paramagnetic to a spin
glass phase, which we approach by looking at the divergence of the
spin relaxation rate at the transition point. In the aging regime, we
determine the dynamical equations governing the time evolution of the
spin response and correlation functions, and show that all terms in
the equations that arise solely from quantum effects are irrelevant at
long times under time reparametrization group (R$p$G)
transformations. At long times, quantum effects enter only through the
renormalization of the parameters in the dynamical equations for the
classical counterpart of the rotor model. Consequently, quantum
effects only modify the out of equilibrium fluctuation dissipation
relation (OEFDR), i.e. the ratio $X$ between the temperature and the
effective temperature, but not the form of the classical OEFDR.
\end{abstract}

] 

\narrowtext
\section{Introduction}
\label{sec:intro}
In recent years the focus of studies of glassy systems has moved from
equilibrium to non-equilibrium properties, reflecting the understanding that
glassiness is an intrinsically non-equilibrium phenomenon.  The effect of
quantum mechanics on spin glasses has also been a topic of interest.  On the
experimental side, there have been several studies of spin systems which are
in the vicinity of a quantum phase transition from a spin-glass to a
spin-fluid \cite{Wu,Keimer,Hayden,Broholm,SY} including the dipolar,
transverse-field Ising magnet Li Ho$_x$ Y$_{1-x}$ F$_4$. \cite{Wu} On the
theoretical side, studies have mainly focussed on the Ising model 
in a transverse
field or quantum rotors.\cite{HuseM,Thill,RozGrem,KopUs,DFisher,Motrunich,YSR,RSY,Dalidovich}  

There has also been much recent work on the quantum-classical connection in
spin glasses, \cite{RSY,CugLoz,MCshort,QTAP,CGSS,Ritort,SUN} such as the
derivation of quantum TAP equations, \cite{QTAP} and the large $N$ solution
of $SU(N)$ models of spin glasses. \cite{SUN} Quantum fluctuations have been
shown to be responsible for redefining the boundary between the spin glass
and the quantum disordered phases in systems such as the rotors or the
transverse field Ising model. In systems such as the quantum extension of the
spherical $p$-spin-glass, quantum fluctuations are responsible for a
crossover from a second to a first order phase transition. \cite{CGSS}
 
Another system in which quantum effects and glassiness coexist is in the stripe
glasses found in doped Mott insulators (e.g. Ref. \onlinecite{Schmalian}).

In principle, quantum effects could also alter the non-equilibrium dynamics
in the glassy regime at very low temperatures.  Whilst the properties near
the $T=0$ quantum critical point of these systems can be understood within a
static formulation, \cite{RSY} studying the non-equilibrium effects in the
glassy system requires a quantum dynamical approach.

Aging effects in classical spin glasses and other glassy systems have
been investigated considerably in recent years. The out of equilibrium
nature of the system displays itself through a persistent dependence
of susceptibilities on the waiting time, the time since the system
entered the glassy phase.  \cite{BCKM} The correlation functions in
such systems often display quite general scaling features,
\cite{Coniglio} and may also have an invariance under time
reparametrizations.  \cite{FM} It is a natural question to ask whether
quantum spin glasses have a similar aging behaviour to their classical
counterparts.

Understanding aging dynamics in quantum systems requires formulating
the problem within a non-equilibrium closed time path (CTP) formalism,
such as the Schwinger-Keldysh approach. This dynamical approach is
particularly attractive for studying disordered systems, since it
eliminates the need to use replicas in carrying out the average over
quenched disorder.  This is because the generating functional in the
CTP formalism is independent of the disorder realization. Several
different disordered quantum systems have been studied recently using
this approach, e.g. disordered and interacting electronic systems,
\cite{Chamon,Andreev} the infinite-range quantum $p$-spin glass in the
spherical limit, \cite{CugLoz} and a semionic representation for
quantum spin systems.  \cite{KislevOppermann}

In this paper, using the Schwinger-Keldysh dynamical approach, we
investigate the aging behaviour of a spin glass of $M$ component
quantum rotors that have infinte range interactions and on site self
interactions with a coefficient $u$. The model of quantum rotors is
considerably simpler than that of true quantum Heisenberg spins
present in any isotropic antiferromagnet or systems like the doped
cuprates; the different components of the rotor variables all commute
with each other, unlike the quantum spins.  As a consequence, the
path-integral written in the rotor variables has an action which
contains no Berry phases and is purely real.

The quantum rotors are particularly interesting in elucidating the
role of quantum mechanics in the long time dynamics of the glassy
phase for a number of reasons.  They provide an example of a quantum
system with continuous replica symmetry breaking (RSB), in contrast to
the one-step RSB quantum $p$-spin model with a spherical constraint,
which was recently studied by Cugliandolo and Lozano. \cite{CugLoz}
Another feature of the rotor system is that it contains an expansion
parameter, $u$, which we can use to organize a perturbative
expansion. When we consider the model to $O(u)$, we get equations that
look very similar to those obtained in the case of the $p=2$ spherical
spin model, which has no RSB. However, when we expand to $O(u^2)$ we
end up with terms that are similar to those from the $p=4$ spherical
model and also some terms that do not arise in $p$-spin
models. Therefore, we can investigate very generally how each term
contributes to the long time dynamics of the model, and how the
quantum effects enter in the problem through each of these terms.

We show that the terms in the dynamical equations that appear as a
consequence of quantum mechanics are irrelevant at long times. The
precise sense in which the terms are irrelevant was defined in
Ref. \onlinecite{MCshort}, using the language of time reparametrization
invariance and the reparametrization group (R$p$G) of time
transformations. Hence, at very long times, the dynamics of the
quantum rotors is completely determined by a {\it renormalized
classical} version of the model. In this time regime, quantum mechanics
enters the problem only through a renormalization of the
coefficients in the dynamical equations, and there is a complete
correspondence between the classical and quantum versions of the model
in the aging regime. In particular, this implies that the response and
correlation functions of the quantum rotors are related by a modified
fluctuation-dissipation relation, with an effective temperature
$T_{\rm eff}>T$, just as found in the classical spin glass models. Our
results within the R$p$G framework extend the connection between the
aging regimes of quantum and classical systems, as found in the
quantum $p$-spin model \cite{CugLoz} with one-step RSB, to a wider and
more general class of quantum glassy models.

The paper is organized as follows. In section \ref{sec:model}, we
describe the model that we study and the CTP formalism that we use.
In section \ref{sec:saddle} we obtain the saddle point dynamical
equations for the model and develop the perturbation theory in $u$
that we need to obtain these equations. In section \ref{sec:para} we
study the solutions to these equations in the paramagnetic phase. In
section \ref{sec:SG} we investigate the solutions to the dynamical
equations in the spin glass phase, and show that a selected number of
terms can be made invariant under reparametrizations of the time
coordinate in the aging regime. We analyze the terms stemming from the
quantum dynamics imposed on the system that are not present in the
classical model, and show that they become irrelevant at long times
under the time reparametrization transformations. Finally, in section
\ref{sec:discussion} we discuss differences between classical and
quantum spin glasses.

\section{The Model}
\label{sec:model}
We consider the model of a glass of quantum rotors introduced in Refs.
\onlinecite{YSR,RSY} with long range
interactions on a $d$-dimensional lattice with $N$ sites. 
(We let $d$ and $N$ go to infinity later).  An
important point to note is that the $M$ components of angular momentum
of the rotors commute on the same site, unlike Heisenberg spins, for
which there are non-trivial commutation relations between the
components.  This simplification allows us to write a path integral in
which there are no Berry phase terms, and makes the rotors much easier
to treat analytically.  

To derive the dynamical equations, we use a CTP (Schwinger-Keldysh)
formalism. \cite{CugLoz,Chamon,Andreev,Schwinger,Keldysh} The CTP
approach provides a way to study the non-equilibrium response of a
system. The price that has to be paid for this is the introduction of
a second component to the system, with time flowing in the opposite
direction.  The non-equilibrium formulation has the advantage that it
leads to a generating functional that is automatically normalized to
unity. The property of normalization allows averages over disorder
realizations in a way that bypasses using replicas.

\begin{figure}[htb]
\centerline{\psfig{file=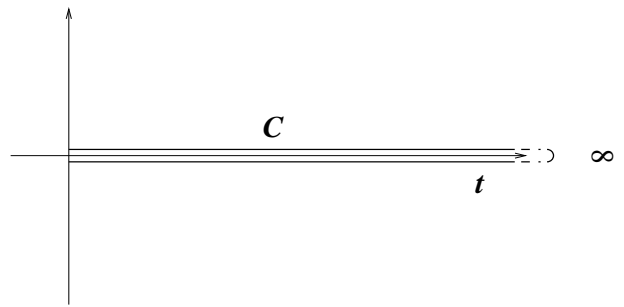,height=4cm,width=8cm,angle=270}}
\caption{The contour ${\mathcal C}$ used in the formalism.}
\label{fig:contour}
\end{figure}

There are many possibilities for the choice of integration contour in
the complex time plane. \cite{Chou,Landsman-Weert}  The contour
${\mathcal C}$ that we work with is illustrated in Fig.
\ref{fig:contour}, which starts at $t=0$, runs to $t=\infty$, and then
returns in the negative $t$ direction to $t=0$. The contours more
usually used, for example in problems involving the calculation of
non-linear response, run from $t = -\infty$ to $t=\infty$ and back
again; \cite{Chou,Landsman-Weert} however after an infinite time, the
system will have equilibrated, so it is no longer possible to study
the non-equilibrium dynamics that we are interested in.

For a system at equilibrium, the usual way to introduce finite
temperature is through the density matrix with the form of the
Gibbs-Boltzmann distribution.  In a non-equilibrium situation, the
system cannot be described by the Gibbs-Boltzmann distribution, so an
alternative approach is required. The solution is to couple the system
to a heat bath \cite{Feynman} (chosen here to be a set of independent
harmonic oscillators) and then allow the system to reach a constant
temperature. A detailed account of integrating the bath variables to
obtain an effective action in the $p$-spin model is given in
Ref. \onlinecite{CugLoz}, and we follow their approach in our study of the
quantum rotors.

\subsection{Effective action for the rotors and bath system}

The action for the system of rotors interacting with the bath takes
the form
\begin{equation}
{S} = {S}_{\rm free} + {S}_{\rm int} 
+ {S}_{\rm dis} + {S}_T \; ,
\end{equation}
where ${S}_{\rm free}$ is the free action, ${S}_{\rm int}$
describes the self-interaction of the rotors, ${S}_{\rm dis}$
contains the spin exchange interactions which introduce disorder and
frustration to the model, and ${S}_T$ describes the
interaction with an external heat bath.  The Lagrangian for the free
rotors is

\begin{equation}
{\mathcal L}_{\rm free} = \sum_{i}
\frac{1}{2g}\;(\partial_t S^a_{i\mu})\;
\sigma_3^{ab}\;
(\partial_t S^b_{i\mu}) + \frac{1}{2}
m^2 \;S^a_{i\mu}\;\sigma_3^{ab}\;S^b_{i\mu} \; .
\end{equation} 
The indices on the spin variable $S^a_{i\mu}$ refer to the Keldysh
contour branch ($a$), site ($i$), and spin component ($\mu$); sums
over repeated Keldysh and spin component indices are always implied
unless explicitly noted otherwise. The $\bvec{\sigma}_{1,2,3}$
are the standard Pauli sigma matrices.  Note that the $S^2$ terms
enter with an opposite sign to the $S^1$ terms because the direction
of time integration is reversed on the second part of the contour. The
first term in the Lagrangian ${\mathcal L}_{\rm free}$ is a kinetic
energy term ($1/g$ is the moment of inertia), and the second is a
potential energy term ($m^2$ also acts as the parameter that tunes
between the paramagnetic and spin glass phases) -- the free Lagrangian
may be thought of as $M$-component harmonic oscillators on a lattice,
each at site $i$.

The self-interactions are described by
\begin{equation}
{\mathcal L}_{\rm int} = \sum_{i} \frac{u}{2}
\left\{\left[(S^1_{i\mu})^2\right]^2 - 
\left[(S^2_{i\mu})^2\right]^2\right\}.
\end{equation}
The strength of the self-interaction is $u$ and this term is necessary
to ensure the stability of the system, since otherwise the Lagrangian
would describe randomly coupled harmonic oscillators. \cite{RSY} 

The term in the Lagrangian that contains quenched disorder is

\begin{equation}
\label{ldis}
{\mathcal L}_{\rm dis} = 
\sum_{i\neq j} J_{ij}\; S^a_{i\mu}\; \sigma_3^{ab}\; S^b_{j\mu}.
\end{equation}
Thermal effects are introduced by placing the system in contact with a
heat bath, using the method introduced by Feynman and
Vernon. \cite{Feynman} The contribution of this coupling to the action
after integrating out heat bath variables is

\begin{eqnarray}
S_T &&=-\int dt_1 dt_2 \sum_i \eta (t_1 - t_2)\; S^a_{i\mu}(t_1)\;
[\sigma_1^{ab} + i\sigma_2^{ab}]\; S^b_{i\mu}(t_2) 
\nonumber \\ 
&&+ i\int dt_1 dt_2 \sum_i
\nu(t_1 - t_2)\; S^a_{i\mu}(t_1)\;[\delta^{ab} - \sigma_1^{ab}]\;
S^b_{i\mu}(t_2),\nonumber \\
&&
\end{eqnarray}
where the noise and dissipative kernels, $\nu$ and $\eta$
respectively, are given by

\begin{eqnarray}
\nu(t)&&= \int_0^\infty d\omega \; I(\omega) 
\coth\left(\frac{\beta\hbar\omega}{2}\right) \; \cos (\omega t), \\
\eta(t) &&= - \theta(t)\int_0^\infty d\omega \; I(\omega) 
\; \sin(\omega t),
\end{eqnarray}
and $I(\omega)$ is the spectral density of the bath

\begin{equation}
I(\omega) = \sum_{n=1}^{N_b}\delta(\omega - \omega_n) 
\frac{C_n^2}{2M_n\omega_n},
\end{equation}
where $N_b$ is the number of oscillators in the bath, $\omega_n$ is
the natural frequency of the $n^{th}$ oscillator, $M_n$ is its mass,
and $C_n$ is its coupling to the system.  Here we only consider the
case of ohmic dissipation
$$I(\omega) =
\frac{\gamma}{\pi}\omega e^{-\omega/\Lambda}, \quad {\rm for} \quad \omega <
\Lambda,$$  
where $\gamma$ plays the role of a friction coefficient (to relate to the notation
in previous works, \cite{CugLoz,MCshort} $\gamma = M\gamma_0$).  

Properties to note are that both kernels are purely real, $\nu(t) =
\nu(-t)$, and both kernels decay rapidly at large time differences.
Note that the quantum fluctuation-dissipation theorem (QFDT) 
holds for the bath variables when there is ohmic
dissipation.  The QFDT relating the heat bath variables is
\cite{CugLoz}

\begin{equation}
\label{eq:QFDT}
\eta(\omega) = 
\frac{1}{\hbar}\lim_{\epsilon\rightarrow 0}\int \frac{d\omega^\prime}{2\pi}
\frac{1}{\omega - \omega^\prime + i\epsilon} \tanh\left(\frac{\beta\hbar\omega^\prime}{2}
\right) \hbar\nu(\omega^\prime).
\end{equation}

\subsection{Disorder average and definition of correlation functions}

Having described all of the terms in the action, we now move on to
performing the average over disorder realizations and defining the 
correlation and response in terms of the CTP two-point correlators.

The closed time path generating functional is

\begin{equation}
Z = \int [DS^a_{i\mu}]\;\; 
e^{\frac{i}{\hbar} S}\;  . 
\end{equation}
We perform an average over the quenched disorder

\begin{equation}
\overline{Z} = \int [DJ] \; {\mathcal P}(J_{ij})\;
\int [DS^a_{i\mu}] \;e^{\frac{i}{\hbar}S},
\end{equation}
where 

\begin{equation}
\label{pij}
{\mathcal P}(J_{ij}) =
\left(\frac{N}{2\pi J^2}\right)^{\frac{1}{2}} \exp\left[
-\frac{NJ_{ij}^2}{2J^2}\right] \; ,
\end{equation}
for a long range interaction.  The form Eq. (\ref{pij}) for the
disorder probability distribution comes from the assumption that the
disorder is Gaussian distributed and that $\overline{J_{ij}^2} =
J^2/N$ .  We can perform the integration over disorder without
replicas because of the normalization property of the generating
functional in the CTP formulation, since the disorder and the initial
conditions are uncorrelated:

\begin{eqnarray}
&& \int [DJ] \; {\mathcal P}(J_{ij}) \;
e^{\frac{i}{\hbar}\int dt 
\sum_{i\neq j}J_{ij}
S^a_{i\mu}(t)\; \sigma_3^{ab}\; S^b_{j\mu}(t)} = 
\left(\frac{N}{2\pi J^2}\right)^{\frac{1}{2}} \nonumber \\
& & e^{-\frac{J^2}{2N\hbar^2}\int 
dt_1 dt_2 \sum_{i\neq j}
\left[ S^a_{i\mu}(t_1)\sigma_3^{ab}S^b_{j\mu}(t_1)
S^c_{j\nu}(t_2)
\sigma_3^{cd} S^d_{i\nu}(t_2)\right]} . \nonumber \\
\end{eqnarray}
The dynamical equations are written in terms of the correlation and response,
which are defined below.  We use an overbar $\overline{\cdots}$ to indicate
an average over realizations of disorder and angular brackets
$\left<\ldots\right>$ to indicate an average with respect to the action. The
correlation, $C(t_1,t_2)$, and response, $R(t_1,t_2)$, are

\begin{equation}
C_{ij,\mu\nu}(t_1,t_2)=\frac{1}{2} \overline{\left<S^1_{i\mu}(t_1)S^1_{j\nu}(t_2) +
S^2_{i\mu}(t_1)S^2_{j\nu}(t_2) \right>},  
\end{equation}
and $R_{ij,\mu\nu}(t_1,t_2) = {\delta
\overline{\left<S_{i\mu}^1(t_1)\right>} / \delta h_{j\nu}(t_2) }$,
which in linear response theory may be written as

\begin{equation}
R_{ij,\mu\nu}(t_1,t_2) =
\frac{i}{\hbar}\overline{\left<S^1_{i\mu}(t_1)[S^1_{j\nu}(t_2)-
S^2_{j\nu}(t_2)]\right>} \; .
\end{equation}

An alternative approach to the dynamics is to 
introduce a matrix propagator as in the Keldysh
formalism, \cite{Keldysh} which takes the form for bosonic fields (e.g. Ref. 
\onlinecite{Rammer})

\begin{eqnarray}
G^{11}_{ij,\mu\nu}(t_1,t_2) & = & -i\left<{\mathcal T}\left[S_{i\mu}^1(t_1)S_{j\nu}^1(t_2)
\right]\right> , \\
G^{12}_{ij,\mu\nu}(t_1,t_2) & = & -i\left<S_{j\nu}^2(t_2)S_{i\mu}^1(t_1)\right> , \\
G^{21}_{ij,\mu\nu}(t_1,t_2) & = & -i\left<S_{i\mu}^2(t_1)S_{j\nu}^1(t_2)\right> , \\
G^{22}_{ij,\mu\nu}(t_1,t_2) & = & -i\left<\tilde{\mathcal T}\left[S_{i\mu}^2
(t_1)S_{j\nu}^2(t_2)
\right]\right> , 
\end{eqnarray}
where ${\mathcal T}$ and $\tilde{\mathcal T}$ are the operators for
time ordering and anti-time ordering respectively.  Under the
transformation in Keldysh space $G \rightarrow \hat{G} = LGL^\dagger$,
where

\begin{equation}
L = \frac{1}{\sqrt{2}}({\mathbf 1} - i \sigma_2),
\end{equation}
the propagator takes the form

\begin{eqnarray}
\hat{G} = 
LGL^\dagger = \left(\matrix{0 & G^A \cr G^R & G^K }
\right),
\end{eqnarray}
where

\begin{eqnarray}
G^R & = & G^{11} - G^{12} = G^{21} - G^{22}, \\
G^A & = & G^{11} - G^{21} = G^{12} - G^{22}, \\
G^K & = & G^{11} + G^{22} = G^{21} + G^{12},
\end{eqnarray}
and the results follow from the definitions of the propagators in terms of 
time ordered products. The definitions above imply the following
properties for the propagators

\begin{eqnarray}
G^R_{ji,\nu\mu}(t_2,t_1) & = & G^A_{ij,\mu\nu}(t_1,t_2) , \\
G^K_{ji,\nu\mu}(t_2,t_1) & = & G^K_{ij,\mu\nu}(t_1,t_2) .
\end{eqnarray}
The retarded and advanced propagators are purely real and the Keldysh 
propagator is purely imaginary. Under the transformation, $L$, the spins are 
also rotated,
$S \rightarrow \tilde{S} = LS$,
so for 
\begin{equation}
S = \left(\matrix{ S^1 \cr S^2
}\right),
\end{equation}
we have 

\begin{equation}
\tilde{S} = \frac{1}{\sqrt{2}}\left(\matrix{
S^1 - S^2 \cr S^1 + S^2 }\right) = \left(\matrix{ \hat{S} \cr S
}\right).
\end{equation}
 The $\hat{S}$ and $S$ variables in the quantum CTP formulation
naturally become the usual variables within the Martin-Siggia-Rose
(MSR) formalism \cite{MSR} when the classical limit is taken. The
relation between the Keldysh and retarded propagators and the
correlation and response is

\begin{eqnarray}
G^K(t_1,t_2) & = & -2i C(t_1,t_2) , \\
G^R(t_1,t_2) & = & -\hbar R(t_1,t_2). 
\end{eqnarray}
We will work with the correlation and response rather than the
propagators in the dynamical equations.  However, it is easier to use
the matrix propagator in the Feynman rules for including interactions,
rather than using a diagrammatic technique where response and
correlation are treated differently. \cite{DeDominicis}

In anticipation of the saddle-point evaluation for the disorder term,
we perform a Hubbard-Stratonovich transformation to decouple the four
spin term generated by the disorder average. Introducing the
Hubbard-Stratonovich field $Q^{ab}_{ij,\mu\nu}(t_1,t_2)$, the
effective action can be rewritten as

\begin{eqnarray}
Z & = & \int [DQ] e^{-\frac{N}{2J^2} \int dt_1 dt_2 \sum _{i} Q^{ba}_{i,\nu\mu}(t_2,t_1)
Q^{ab}_{i,\mu\nu}(t_1,t_2)} Z[Q],  \\
Z[Q] & = & \int [D\tilde{S}] e^{\frac{i}{\hbar}S_{\rm free} + 
\frac{i}{\hbar}S_Q + \frac{i}{\hbar}S_{\rm int} +
\frac{i}{\hbar} S_T},
\end{eqnarray}
where

\begin{eqnarray}
S_{\rm free} & = & \int dt \sum_i \tilde{S}^a_{i\mu}(t)\;
\Gamma^{ab}_{\mu\nu} \; \tilde{S}^b_{i\nu}(t) \; ,  \\
S_Q & = &  \int 
dt_1 dt_2 \sum_i  Q^{ab}_{i,\mu\nu}(t_1,t_2)\;\tilde{S}^a_{i\mu}(t_1)\;
\sigma_1^{bc} \;\tilde{S}^c_{i\nu}(t_2) \; .
\end{eqnarray}

Next we introduce a modified version of the Hubbard-Stratonovich field,
$\tilde{Q}^{ab}_{i,\mu\nu} = Q^{ac}_{i,\mu\nu} \; \sigma_1^{cb}$, thus

\begin{eqnarray}
Z & = & \int [D\tilde{Q}] e^{-\frac{N}{2J^2} \int dt_1 dt_2 \sum _{i} 
\sigma_1^{a^\prime a}\; \sigma_1^{bb^\prime} \;
\tilde{Q}^{b^\prime a^\prime}_{i,\nu\mu}(t_2,t_1)\;
\tilde{Q}^{ab}_{i,\mu\nu}(t_1,t_2)} \nonumber \\
& & \times \int [D\tilde{S}] e^{\frac{i}{\hbar}S_{\rm free} + \frac{i}{\hbar}
S_{\tilde{Q}} + \frac{i}{\hbar}S_{\rm int} +\frac{i}{\hbar} S_T},
\end{eqnarray}
where

\begin{eqnarray}
S_{\tilde{Q}} = \int 
dt_1 dt_2 \sum_i  \tilde{Q}^{ab}_{i,\mu\nu}(t_1,t_2) \;
\tilde{S}^a_{i\mu}(t_1) \;
\tilde{S}^b_{i\nu}(t_2) .
\label{eq:SQ}
\end{eqnarray}

The $\tilde{\Gamma}^{ab}_{\mu\nu}$ term has 
the structure 
\begin{eqnarray}
\tilde{\Gamma}^{ab}_{\mu\nu} = \delta_{\mu\nu}
\left(\matrix{ \Gamma_K & \Gamma_R \cr
\Gamma_A & 0 } \right),
\end{eqnarray}
where 
\begin{equation}
\Gamma_R = \Gamma_A = -\frac{1}{2g}\frac{\partial^2}{\partial t^2} + 
\frac{1}{2}m^2, \quad \Gamma_K = 0.
\end{equation}

\section{Dynamical equations}
\label{sec:saddle}
Using propagators, the self-consistent mean field equations can be
represented in the form

\begin{equation}
G^{-1} = G_0^{-1} - \Sigma_J - \Sigma_T - \Sigma_u,
\label{eq:SC}
\end{equation}
where the $\Sigma$ terms are the self energies from the disorder, the
interaction with the thermal bath and from self-interaction
respectively. The general strategy that we will adopt in obtaining the
dynamical equations for this model is to perform a saddle point
evaluation of $\Sigma_J$ and then a perturbation expansion in $u$.  We
obtain a solution to $O(u)$ in the paramagnetic phase, in analogy with
the replica symmetric solution in the equilibrium problem, \cite{RSY}
whilst in the spin glass phase we need to consider terms to $O(u^2)$
in the interaction, which are the terms found to contribute to RSB.

\subsection{Saddle point evaluation of $\Sigma_J$}
At the saddle point the functional derivative with respect to the
Hubbard-Stratonovich field is zero
$$\frac{\delta Z}{\delta Q} = 0.$$ 
The variation leads to the following equations

\begin{eqnarray}
0 & = & \frac{N\hbar i}{2J^2} \left<
\sigma_1^{bb^\prime}\;
\tilde{Q}^{b^\prime a^\prime}_{\nu\mu}(t_2,t_1) \;
\sigma_1^{a^\prime a} \right>
\nonumber \\
& & + \sum_i \left<\tilde{S}^a_{i\mu}(t_1)\tilde{S}^b_{i\nu}(t_2)\right> .
\end{eqnarray}
Rearranging gives

\begin{equation}
\left<\tilde{Q}^{ab}_{\mu\nu}(t_1,t_2)\right> = -\frac{2J^2}{\hbar} \;
\sigma_1^{b b^\prime} \;
G^{b^\prime a^\prime}_{\nu\mu}(t_2,t_1) \;
\sigma_1^{a a^\prime}.
\end{equation}
By substituting the saddle solution into Eq. (\ref{eq:SQ}), one
obtains the self-energy contribution coming from the disorder:
\begin{eqnarray}
\Sigma_J(t_1,t_2)=&&
\frac{2J^2}{\hbar} \;
\sigma_1 \; G(t_1,t_2) \;\sigma_1 
\nonumber\\
=&&
\frac{2J^2}{\hbar} \;
\left(\matrix{ G^K & G^R \cr G^A & 0 }
\right)_{(t_1,t_2)} .
\end{eqnarray}
The notation of a matrix with $(t_1,t_2)$ following it is
used to indicate that all of the propagators in the matrix have that
as their argument. At the mean field level the solution is homogeneous
in the site index, so we drop it. In the absence of a magnetic field,
we also drop the spin indicies, i.e. $G_{\mu\nu} = G\;\delta_{\mu\nu}$.

\subsection{Diagrammatic perturbation in $u$}
\begin{figure}[htb]
\centerline{\psfig{file=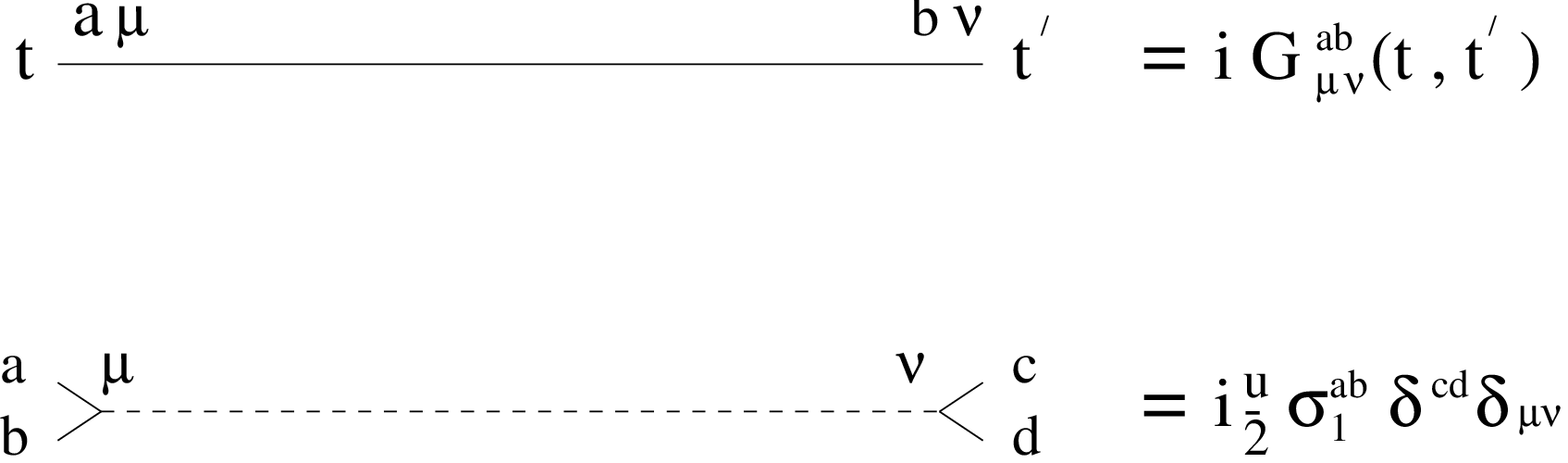,height=2cm,width=8cm,angle=270}}
\caption{The Feynman rules for the pertubation theory in $u$.}
\label{fig:rules}
\end{figure}

The interaction terms in the action may be treated in a diagrammatic way
by noting that in the $\tilde{S}$ basis, the interaction term is written as

\begin{equation}
S_{\rm int} = \frac{u}{2}\int_0^\infty dt \sum_i \tilde{S}^a_{i\mu}(t)\tilde{S}^a_{i\nu}(t)
\tilde{S}^c_{i\mu}(t)\sigma_1^{cd}\tilde{S}^d_{i\nu}(t).
\end{equation}
The Feynman rules (see Fig. \ref{fig:rules}) for calculating
$\Sigma_u$ term by term in $u$ are that the propagator (solid line) is
$iG^{ab}_{\mu\nu}(t,t^\prime)$ and the interaction propagator (dashed
line with a dot at one end) is
$i\frac{u}{2}\sigma_1^{ab}\delta^{cd}\delta_{\mu\nu}$. 
The $\sigma_1$ is inserted
at the end with the dot.

\begin{figure}[htb]
\centerline{\psfig{file=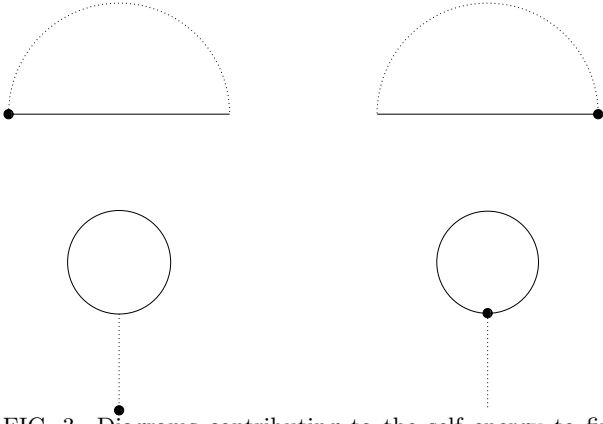,height=5.5cm,angle=270}}
\caption{Diagrams contributing to the self energy to first order in $u$.}
\label{fig:u1}
\end{figure}

This leads to four contributions to the self energy $\Sigma_u$ to
first order in $u$, as shown in Fig. \ref{fig:u1}.  Note that the
bottom two diagrams in Fig. \ref{fig:u1} are proportional to $M$, 
the number of spin components
of the rotors, since there is a trace around a closed loop, and that
the trace includes spin indicies.  Note also that $tr(G\sigma_1) = 0$,
since $G^R(t,t) = 0$.  The sum of these contributions is

\begin{eqnarray}
\Sigma^{(1)}_u(t,t^\prime) & = & 
\frac{iu}{2}(1+M)\;\;\sigma_1\; G^K(t,t) \;\delta(t-t^\prime) \; ,
\end{eqnarray}
where the superscript (1) indicates that the self energy is to first
order in $u$.

\begin{figure}[htb]
\centerline{\psfig{file=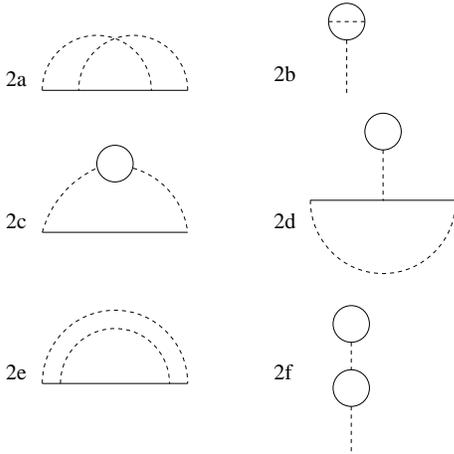,height=6cm,angle=270}}
\caption{Diagrams contributing to the self energy to second order in $u$.}
\label{fig:u2}
\end{figure}

To second order in $u$ there are six different one particle
 irreducible diagram topologies to consider which contribute to the
 self energy and four different diagrams for each topology (the
 different topologies are
 shown in Fig. \ref{fig:u2}).  We treat each topology separately,
 starting with the propagator with two self interactions, which
 contributes

\begin{eqnarray}
\label{self2a}
&& \Sigma_u^{(2a)}(t,t^\prime)= -\frac{u^2}{8\hbar} \times  \\
&& \left(\matrix{ 
[3(G^A)^2 + 3(G^R)^2 + (G^K)^2] G^K & [3(G^K)^2 + (G^A)^2]G^A \cr
[3(G^K)^2 + (G^R)^2]G^R & 0 
} \right)_{(t,t^\prime)} . \nonumber 
\end{eqnarray}

The next diagrams are the vertex correction to the closed loop that
contributes to $\Sigma^{(1)}_u$ and the propagator with one self
interaction, which contribute

\begin{eqnarray}
\label{self2b}
\Sigma_u^{(2b)}(t,t^\prime) & = & -\frac{ u^2}{4\hbar}\;\;\sigma_1\;
\delta(t-t^\prime) \nonumber \\
& & \times \left[
\int_0^\infty d\tilde{t} \, G^K(\tilde{t},\tilde{t})
 G^R(t,\tilde{t})G^K(t,\tilde{t})
\right]  . 
\end{eqnarray}
The remainder of the self energies are expressible in terms of the
expressions found in equations (\ref{self2a}) and (\ref{self2b}):
$\Sigma^{(2c)} = M \Sigma^{(2a)}$, $\Sigma^{(2d)} = M \Sigma^{(2b)}$,
 $\Sigma^{(2e)} = \Sigma^{(2b)}$ and $\Sigma^{(2f)} =
\Sigma^{(2d)}$.  The diagram that plays the most important role in the
physics is (2a) because it depends on two times rather than only one.

\subsection{The dynamical equations}
From the self consistent Eq. (\ref{eq:SC}) we obtain the dynamical
equation
\begin{equation}
{\mathbf 1}= \left(G_0^{-1} - \Sigma_J - \Sigma_T - \Sigma_u\right)\;G
\; .
\end{equation}
By substituting the self-energies $\Sigma_J$, $\Sigma_T$ and $\Sigma_u$ that we
determined above to order $u^2$, and shifting notation to response and
correlation, we obtain the dynamical equations for the quantum rotor
model

\begin{eqnarray}
\label{eq:rsaddle}
\delta(t_1 - t_2) & = & \left\{\frac{1}{2g}\frac{\partial^2}{\partial t_1^2} - \frac{1}{2}m^2
-u(1+M) C(t_1,t_1) \right. \nonumber \\
& &  - 2(1+M) u^2 C(t_1,t_1) \nonumber \\
& & \left. \hspace*{1cm} \times \int_0^\infty dt \, R(t_1,t)C(t_1,t) \right\} R(t_1,t_2)
\nonumber \\
& & + \int_0^\infty dt \, \eta(t_1 - t) R(t,t_2) \nonumber \\
& &  - 2J^2 \int_0^\infty dt \, 
R(t_1,t) R(t,t_2) \nonumber \\
& & - \frac{3}{2}(1+M) u^2 \int_0^\infty dt \, C(t_1,t)^2 R(t_1,t) R(t,t_2) \nonumber \\
& & + \frac{1}{2}(1+M)u^2 \hbar^2 \int_0^\infty dt \, R(t_1,t)^3 R(t,t_2),
\end{eqnarray}

and
\begin{eqnarray}
\label{eq:csaddle}
0 & = & \left\{\frac{1}{2g}\frac{\partial^2}{\partial t_1^2} - \frac{1}{2}m^2
-u(1+M) C(t_1,t_1) \right. \nonumber \\
& & \left. - 2(1+M) u^2 C(t_1,t_1)\int_0^\infty dt \, R(t_1,t)C(t_1,t) \right\} C(t_1,t_2)
\nonumber \\
& & + \int_0^\infty dt \, \eta(t_1 - t) C(t,t_2) - \frac{\hbar}{2} \int_0^\infty 
dt \, \nu(t_1 - t) R(t_2,t) \nonumber \\
& & - 2J^2 \int_0^\infty dt \, \left[C(t_1,t)R(t_2,t) + R(t_1,t)C(t,t_2)\right] \nonumber \\
& & - \frac{1}{2} (1+M) u^2 \int_0^\infty dt \, C(t_1,t)^3 R(t_2,t) \nonumber \\
& & - \frac{3}{2} (1+M) u^2 \int_0^\infty dt \, C(t_1,t)^2 R(t_1,t) C(t,t_2) \nonumber \\
& & + \frac{1}{2} (1+M) u^2 \hbar^2 \int_0^\infty dt \, \left[ 
 R(t_1,t)^3 C(t,t_2) \right. \nonumber \\
& & \hspace*{1cm} + \left. 3R(t_1,t)^2 R(t_2,t)C(t_1,t) \right].
\end{eqnarray}
The terms that arise in this expansion look partly like a $p=4$ spherical spin 
glass (the $C^3R$ term in the correlation equation), but there are also terms 
(multiplied by $\hbar^2$) which do not appear in the classical spherical model.

\section{Paramagnetic phase}
\label{sec:para}
In the paramagnetic phase the correlation and response are time translation
invariant (TTI) after the initial transients have died out. We find a
solution to the dynamical equations (\ref{eq:rsaddle}) and (\ref{eq:csaddle})
to $O(u)$ in this phase. 
With the assumption of TTI correlators, let $\tau = t_1 - t_2$, and then let
$t_1 \to \infty$.  Define $C_\infty = \lim_{t_1 \to \infty} C(t_1,t_1)$,
which is the equilibrium limit of the equal time correlation (we also use
$C_\infty$ in the spin glass phase; as a one time quantity the equal time
correlation has a limit as $t_1 \to \infty$).  TTI allows us to solve the
problem in the paramagnetic phase by performing a Fourier transformation:

\begin{eqnarray}
\label{eq:Rft}
-1 & = & \left\{\frac{\omega^2}{2g} + \frac{1}{2}m^2 + 
u(1+M)C_\infty \right\}R(\omega) 
\nonumber \\ 
& & + i \int_{-\infty}^\infty \frac{d\omega^\prime}{2\pi} 
\frac{R(\omega^\prime)
\eta(\omega^\prime)}{\omega - \omega^\prime + i0} \nonumber \\
& & - 2J^2 i \int_{-\infty}^\infty
\frac{d\omega^\prime}{2\pi} \frac{R(\omega^\prime)
R(\omega^\prime)}{\omega - \omega^\prime + i0} ,
\end{eqnarray}
and

\begin{eqnarray}
\label{eq:Cft}
0 & = & \left\{\frac{\omega^2}{2g} + \frac{1}{2}m^2 + 
u(1+M) C_\infty \right\}C(\omega)
\nonumber \\
& & + i \int_{-\infty}^\infty \frac{d\omega^\prime}{2\pi} 
\frac{\eta(\omega^\prime)
C(\omega^\prime)}{\omega - \omega^\prime + i0} \nonumber \\
& & - \frac{\hbar}{2} i \int_{-\infty}^\infty 
\frac{d\omega^\prime}{2\pi} \frac{\nu(\omega^\prime)
R(-\omega^\prime)}{\omega - \omega^\prime
+ i0} \nonumber \\
& & - 2J^2 i \int_{-\infty}^\infty \frac{d\omega^\prime}{2\pi} 
\left[\frac{C(\omega^\prime)R(-\omega^\prime) + 
R(\omega^\prime)C(\omega^\prime)}{\omega 
- \omega^\prime + i0} \right] .
\end{eqnarray} 

The following properties hold for the correlation and reponse and their
Fourier transforms: $C(t)$ is invariant under time reversal and is real,
hence $C(\omega) = C^*(-\omega)$, $C^*(\omega) = C(\omega)$ and $C(\omega)$
is real.  The response is real, hence $R(\omega) + R(-\omega) = 2 \, {\rm
  Re}[R(\omega)]$ and $R(\omega) - R(-\omega) = 2 i\, {\rm Im}[R(\omega)]$.
In addition, the noise kernel $\nu(\omega)$ is real and has the property
$\nu(\omega) = \nu(-\omega)$.  The real part of the equation (\ref{eq:Cft})
combined with the QFDT for the heat bath variables (\ref{eq:QFDT}) gives

\begin{equation}
\label{rcqfdt}
{\rm Im}[R(\omega)] = \frac{1}{\hbar}\tanh\left(\frac{\beta\hbar\omega}{2}\right)C(\omega).
\end{equation}
Hence the correlation and response are related by the QFDT if the same is
true for the heat bath variables.  In this regime, we can solve explicity for
the response, and hence the correlation through the QFDT (\ref{rcqfdt}). We
use these solutions to investigate the self-consistency condition in the
paramagnetic regime and investigate critical slowing down as discussed by
Sompolinsky. \cite{Somp}

\subsection{Response, correlation, and phase boundary}

Assuming that the integrands in the resonant integrals in equations
(\ref{eq:Rft}) and (\ref{eq:Cft}) fall off sufficiently quickly at
large $\omega$, then they may be rewritten and solved for $R(\omega)$
to give

\begin{equation}
R(\omega) = -\frac{x(\omega)}{4J^2} - \frac{1}{4J^2}\sqrt{x(\omega)^2 - 8J^2},
\end{equation}
where 
$$
x(\omega) = \frac{\omega^2}{2g} + \frac{1}{2}m^2 + u(1+M) C_\infty -
\eta(\omega),
$$
(note that $\eta(\omega)$ has both real and imaginary parts).
In the limit that $\omega \rightarrow 0$, the real part of the kernel $\eta$
is proportional to the frequency cutoff, $\Lambda$, so to remove this, define
$m_1^2 = m^2 - 2 \eta(0)$.  Then at zero frequency at the critical point,

\begin{equation}
\label{eq:critical}
\frac{1}{2}m_1^2 + u (1+M) C_\infty  = 2\sqrt{2}\, J,
\end{equation}
and the imaginary part of the response is

\begin{eqnarray}
\label{eq:imrw}
{\rm Im}[R(\omega)] & = & -\frac{{\rm Im}[\eta(\omega)]}{4J^2} \nonumber \\
& & - \frac{1}{4J^2}  \sqrt{2 \, \left|{\rm Im}[\eta(\omega)]\right|
\sqrt{{\rm Im}[\eta(\omega)]^2 
+  8J^2}} \; .
\end{eqnarray}
To get the self-consistency condition for $C_\infty$ at the critical
point, 
we use equation (\ref{eq:imrw}) and the QFDT

\begin{equation}
C_\infty = \hbar\int_{-\infty}^\infty \frac{d\omega}{2\pi} 
\coth\left(\frac{\beta\hbar\omega}{2}\right)
{\rm Im}[R(\omega)].
\end{equation}
Using
\begin{equation}
{\rm Im}[\eta(\omega)] = -\frac{\gamma\omega}{2}e^{-|\omega|/\Lambda}
\theta(\Lambda - |\omega|),
\end{equation}
and noting that the second integral vanishes, since the integrand is odd, we
get the result 

\begin{equation}
C_\infty  = \frac{\pi \gamma}{48\hbar\beta^2 J^2} + O(\Lambda^2) .
\end{equation}
The quantity $C_\infty$ fixes the equal-time correlation,
and provides a constraint for the rotor size in this model, since we
do not impose a spherical constraint. \cite{CugLoz,CugKurA} In the
spin glass phase $C(t,t)$ should relax to an equilibrium value, $q$,
the Edwards-Anderson order parameter. \cite{EA} This is because it is
a one time quantity -- the same is not true for two time quantities.
The quantity $m_1^2$ acts to ``tune'' the system through the spin
glass transition.  Using equation (\ref{eq:critical}) and separating
$m_1^2$ into a piece that depends on the cutoff and a temperature
dependent piece, we have an expression analagous to the one found for
$r_C(T)$ in the equilibrium study of quantum rotors \cite{RSY}

\begin{equation}
m_1^2(\beta) = m_1^2 - \frac{u(1+M)\gamma}{24\hbar\beta^2 J^2}.
\end{equation}

\subsection{Critical slowing down}
The dynamic transition from the paramagnetic to the spin glass phase was
investigated by Sompolinsky and Zippelius. \cite{SompZ} However, in their
treatment they took the
infinite time limit before, rather than after the $N\to \infty$ limit, the
opposite order of limits than that considered here.  This leads to finite
energy barriers between traps and hence an infinite hierachy of time scales
and an ergodic solution.  In the situtation we consider the system is
confined to a single ergodic component and we are studying the relaxation
within a trap.  \cite{CugKurB}

To connect with the work of Sompolinsky and Zippelius, define 

\begin{eqnarray}
\Gamma^{-1}(\omega) = \frac{\partial}{\partial\omega}R^{-1}(\omega),
\end{eqnarray}
which leads to $\Gamma(\omega) \simeq 1/\omega$ at small frequencies and 

\begin{equation}
\Gamma^{-1}(\omega) = \frac{ 
\frac{\partial\eta}{\partial\omega} - 
\frac{\omega}{g}}{\left(1 - 2J^2 R(\omega)^2\right)},
\end{equation}
which is analagous to their result.

\subsection{Alternative approach -- Schwinger-Keldysh Landau Theory}

In previous work on quantum rotor systems, \cite{YSR,RSY} the approach taken
has been to obtain a replicated field theory in the Hubbard-Stratonovich $Q$
fields. In this and related work, \cite{MCshort} we have obtained directly
dynamical equations for spin correlation functions. In the following we will
briefly show how to set up a Landau theory for the fields $Q$ within a
completely dynamical approach, and connect to both the replica results of
Ref. \onlinecite{YSR,RSY} and the results obtained in the preceeding sections.

In the dynamical Schwinger-Keldysh approach, if the system does reach
equilibrium with the bath (as is the case in the paramagnetic phase), then
the temperature of the heat bath can be introduced through the use of a
density matrix rather than with the heat bath kernels referred to in the
previous section.

Just as in section \ref{sec:model}, integrating over disorder and
introducing a Hermitian Hubbard-Stratonovich field leads to the following
action 

\begin{eqnarray}
Z = \int [D\tilde{Q}] e^{-\frac{1}{2J^2}\int dt_1 dt_2 \sum_{ij} \tilde{Q}_i
\sigma_1 K^{-1}_{ij} \tilde{Q}_j \sigma_1 - \log Z_0[\tilde{Q}]},
\end{eqnarray}
where $Z_0$ is the single site generating functional 

\begin{eqnarray}
     Z_{0}[\tilde{Q}] && = \int {\cal D} S^{a} \exp [i S_{0}] , \nonumber  \\
     S_{0} && =  S_{\rm free} +S_{\rm nl} \\
     && - \int dt d t^{\prime} \sum_{i} 
     \tilde{Q}^{ab}_{ i \mu \nu}(t,t^{\prime}) 
     S^{b}_{i \nu}(t^{\prime}) S^{a}_{ i \mu}(t)   \; ,
\nonumber 
\end{eqnarray}
where $S_{\rm nl}$ includes the self-interactions and interaction with
the bath.  We can expand $Z_S[\tilde{Q}]$ by obtaining the
perturbative vertices in powers of the $\tilde{Q}$ fields, noting that
the mean-field solution is isotropic in space. Using standard
diagrammatic methods and in simplified notation, the effective action
up to the cubic order is

\begin{eqnarray}
   S & = & \int \frac{d\omega}{2\pi} Tr( Q G_{0} )
 + \int \frac{d\omega_1}{2\pi}\;
\frac{d\omega_2}{2\pi}\; Tr ( Q G_{0} )^{2} \nonumber \\
 & & + \int \frac{d\omega_1}{2\pi}\;
\frac{d\omega_2}{2\pi}\;
\frac{d\omega_3}{2\pi}\; Tr ( Q G_{0} )^{3}   \nonumber  \\
 & &  - i u \int \frac{d\omega_1}{2\pi}\;
\frac{d\omega_2}{2\pi}\;
\frac{d\omega_3}{2\pi}\;
   Tr [ (G_{0} Q G_{0}) \sigma_{1} (G_{0} Q G_{0} )] \nonumber \\
& & -i u \int \frac{d\omega_1}{2\pi}\;
\frac{d\omega_2}{2\pi}\;
\frac{d\omega_3}{2\pi}\;
   Tr (G_{0} Q G_{0}) \nonumber \\
& & Tr( \sigma_{1} G_{0} Q G_{0} )    
    + \ldots  ,
\label{action}
\end{eqnarray}
where $ \ldots $ means higher order terms.  It is important to point
out that terms with coefficient $u$ are important to get stable saddle
point solutions.  We study the saddle point solutions in the absence of 
spatial fluctuations.

Consider the following saddle point solution which is $ O(M) $ and TTI:
\begin{eqnarray}
Q^{ab}_{\mu\nu}(\omega_1,\omega_2) & = & 
2\pi\;\delta(\omega_1+\omega_2)\;
\delta_{\mu\nu}\;
Q^{ab}(\omega_1)\ , \\
     Q & = & \left( \matrix{
                                Q_{K}  &   Q_{R}  \cr
                                Q_{A}  &   0  \cr
                                }   \right ) .
\label{sp}
\end{eqnarray}
This Ansatz has the same structure in frequency and $ O(M) $ indices
as in the replica approach. \cite{RSY} However, here we are dealing
with a $ 2 \times 2 $ matrix in Keldysh space in contrast to the $ n
\times n $ ( $ n \rightarrow 0 $ limit ) replica matrix.  This leads
to the saddle point action

\begin{eqnarray}
 S_{\rm sp} & = &  \int \frac{d\omega}{2\pi}\;
   ( Q_{R} G^{0}_{R} + Q_{A} G^{0}_{A} )
    \nonumber \\
    & & + \int \frac{d\omega}{2\pi}\;
  [ ( Q_{R} G^{0}_{R})^{2} + ( Q_{A} G^{0}_{A} )^{2} ] \nonumber \\
 & &  + \int \frac{d\omega}{2\pi}\;
  [ ( Q_{R} G^{0}_{R})^{3} + ( Q_{A} G^{0}_{A} )^{3} ]  \nonumber   \\
 & - & iu (1 + M) \int \frac{d\omega}{2\pi}\;
  [  Q_{R} (G^{0}_{R})^{2} +  Q_{A} ( G^{0}_{A} )^{2} ] \nonumber \\
& &  \times \int \frac{d\omega}{2\pi}\;
  [ ( Q_{R} G^{0}_{R} +  Q_{A} G^{0}_{A} ) G^{0}_{K} + Q_{K} G^{0}_{R}
    G^{0}_{A} ]   \nonumber  \\
 & + &  \ldots \quad .
\end{eqnarray}
By using the analytic properties of retarded and advanced Green functions, 
we can remove the quadratic term  by a shift $ Q(\omega) \rightarrow 
Q(\omega)- C \Gamma(\omega) $.  In the low energy limit, using free propagators:
   $ G^{0}_{R} \sim G^{0}_{A} \sim \omega^{2} -r,
  \; G^{0}_{K} \sim 0 $, we get:
\begin{eqnarray}
 S_{\rm sp} & = &  \int \frac{d\omega}{2\pi}\;
   ( \omega^{2} -r )( Q_{R} + Q_{A} )
   + \frac{\kappa}{3} \int \frac{d\omega}{2\pi}\;
  [  Q^{3}_{R}  +  Q^{3}_{A}  ]  \nonumber   \\
 & - & i u(1 + M) \int \frac{d\omega}{2\pi}\;
  [  Q_{R} +  Q_{A} ] \int \frac{d\omega}{2\pi}\; Q_{K} 
   +\ldots \quad .
\label{stat}
\end{eqnarray}

\subsubsection*{Solution in the Paramagnetic phase}

Minimizing Equation (\ref{stat}) with respect to $ Q_{R} $ and $ Q_{A} $, gives
\begin{eqnarray}
  \omega^{2}-r + Q^{2}_{R}-iu (1+M) \int \frac{d\omega}{2\pi}\; Q_{K} 
 =0  ,  \\
  \omega^{2}-r + Q^{2}_{A}-iu (1 +M) \int \frac{d\omega}{2\pi}\; Q_{K} 
 =0  .
\label{para}
\end{eqnarray}
The two equations are complex conjugates of each other, therefore there is only one
independent equation.  Solving  for $ Q_{R} $ gives

\begin{eqnarray}
    Q^{2}_{R} & = & -( \omega^{2} - \tilde{r} )  ,  \\
    \tilde{r} & = &  r - u(1+M) \int \frac{d\omega}{2\pi}\;
   \coth \frac{\beta \omega}{2} \chi^{\prime \prime }( \omega ), \\  
    \chi^{\prime \prime}( \omega) & = & 
     {\rm sgn}( \omega) ( \omega^{2} - \tilde{r} )^{1/2} \theta( \omega -
     \sqrt{ \tilde{r} }  )   .
\end{eqnarray}

The critical point is located at $ \tilde{r} =0 $ where  $ r=r_{c}(T)
= u(1+M) \int \frac{d\omega}{2\pi}\; \omega \; 
\coth \frac{\beta \omega}{2} $.  These results are the same as 
those obtained by the replica approach in Ref.\onlinecite{RSY}.

\section{Glassy Phase}
\label{sec:SG}

In the spin glass phase the correlation and response are no longer
time translation invariant. Instead, there is an aging piece that
remains a function of two times. Below we
investigate the dynamical behavior of the correlation and response
functions for the quantum rotors in the glassy phase.

\subsection{Weak ergodicity breaking and weak long term memory}
Weak ergodicity breaking (WEB) and weak long term memory (WLTM) are phenomena
that have been observed in the solutions to classical mean field spin glass 
models. \cite{CugKurA,Bouchaud,CugKurC}  The numerical results for the 
quantum version of the $p$-spin model \cite{CugLoz} are in agreement with 
those predicted from these scenarios, and we also assume WEB and WLTM 
in our solution of the quantum rotor model.

WEB can be summarized as follows: the correlation function behaves as the
sum of a TTI piece and a piece which depends on two times

\begin{equation}
\label{eq:csplit}
C(t_1,t_2)  =  C_{ST}(t_1 - t_2) + C_{AG}(t_1,t_2) \; .
\end{equation}
The stationary (TTI) piece decays to zero as $t_1 - t_2 \to \infty$, and the 
aging piece satisfies $C_{AG}(t,t) = q$, and 

\begin{equation}
\lim_{t_1 \to \infty} C_{AG}(t_1,t_2) = 0,
\end{equation}
so that both times are important in the decay of the aging piece of 
the correlation.  We separate the response in a similar way to the correlation

\begin{eqnarray}
\label{eq:rsplit}
R(t_1,t_2) & = & R_{ST}(t_1 - t_2) + R_{AG}(t_1,t_2) \; , 
\end{eqnarray}
The stationary piece decays to zero as $t_1 - t_2 \to \infty$, whilst the
decay of the aging piece depends on both times. 
The assumption of WLTM applies to the integral of the response
function: the integral of the response over any finite time 
interval vanishes, i.e.

\begin{equation}
\lim_{t_1 \to \infty} \int_0^t dt_2 \, R(t_1,t_2) = 0 \; ,
\end{equation} 
for fixed $t$, however the integral over an interval that grows with time 
is finite

\begin{equation}
\lim_{t_1 \to \infty} \int_0^{t_1} dt_2 \, R(t_1,t_2) \neq 0 \; .
\end{equation}

\subsection{Modified QFDT} 
Treating the spin glass regime in a similar way to the 
paramagnetic phase, and looking in the
limit $t_1 \to \infty$, it is possible to get an equation 
that includes the TTI pieces, and 
also some static pieces due to a non-zero Edwards-Anderson order parameter.    

In the spin glass phase, we need to take into account the aging that
has occurred up until the waiting time.  \cite{CugLoz,CugKurA} To do
this we split $R$ and $C$ into stationary and aging pieces before
letting $t_1 \to \infty$ and then Fourier transforming.  The relation
we find between the stationary parts of correlation and response is

\begin{eqnarray}
\label{eq:modqfdt}
{\rm Im}[R_{ST}(\omega)] & =&
\frac{1}{\hbar}\tanh\left(\frac{\beta\hbar\omega}{2}\right)C_{ST}
(\omega) \nonumber \\
& & +
\frac{2\pi\beta J^2}{\gamma}q \left(\chi_\infty + 
(RC)_\infty \right) \delta(\omega) \; ,
\end{eqnarray}
where 

\begin{equation}
\chi_\infty = \lim_{t_1 \to \infty} \int_0^{t_1} dt \, R_{ST}(t_1 - t) \; ,
\end{equation}
and 
$(RC)_\infty$ is (noting that $t_1 \sim t_2 \gg |t_1 - t_2|$ in the 
evaluation of the integral)

\begin{equation}
(RC)_\infty = \lim_{t_1 \to \infty} \int_0^{t_1} dt \, R_{AG}(t_1,t)\left(C_{AG}(t,t_1)
+ C_{AG}(t_1,t)\right) \; .
\end{equation}
Equation (\ref{eq:modqfdt}) shows how the QFDT is modified in the 
spin glass phase for the rotor model.  There is the usual piece that is 
present in the paramagnetic phase, and then a delta function at zero frequency
that grows with decreasing temperature.

\subsubsection*{Modified QFDT within the dynamical Landau theory}

As in the treatment above, in the spin glass state, the Keldysh
component $ Q_{K} $ acquires a non-trivial $ \delta $ function part
due to a nonzero Edwards-Anderson order parameter
\begin{equation}
    iQ_{K} =i Q^{reg}_{K} +2 \pi q \delta(\omega) .
\label{delta}
\end{equation}
Substituting the ansatz  into Equation (\ref{para}) gives
\begin{equation}
  \omega^{2}-r + Q^{2}_{R}-i u(1+M) \int \frac{d\omega}{2\pi}\; Q^{reg}_{K}
   -(u+Mv) q =0,
\end{equation}
and solving for $ Q_{R} $ and $ q $ leads to
\begin{eqnarray}
    Q_{R} & = & -i \omega  ,   \\
    q & = & \frac{1}{ u(1+M)} ( r_{c}(T) -r ).
\end{eqnarray}
Again, these results are the same as those obtained by the replica 
approach in Ref.\onlinecite{RSY}.

\subsection{Aging regime}
Aging behaviour has been shown to be a feature in the long time dynamics of
many glassy systems, both in experiments \cite{Svedlindh,Struick} and
theoretical models (see Ref. \onlinecite{BCKM} for a review). In the case of spin
glasses, the major models that have been studied in the aging regime are the
Sherrington Kirkpatrick \cite{SK} (SK) model,\cite{CugKurB} the spherical
$p$-spin model, \cite{CugKurA} and the quantum version of the spherical
$p$-spin model.  \cite{CugLoz} In our work we consider the quantum version of
the rotor model, which in the absence of a magnetic field, is similar to the
soft spin SK model with quantum dynamics rather than Langevin dynamics,
having $M$ components instead of only one.

One of the features of equilibrium solutions of mean field spin glass models
is replica symmetry breaking, introduced by Parisi \cite{Parisi} in the
context of the SK model.  Cugliandolo and Kurchan \cite{CugKurB} started from
a set of equations for the long time dynamics of the SK model and showed that
the correlations have an ultrametric structure analagous to that found in the
Parisi RSB solution of the SK model.  This ultrametric structure in the long
time dynamics can be obtained in the model here only by including terms to
$O(u^2)$ in the dynamical equations -- there is a direct analogy with the
equilibrium solution of quantum rotors, \cite{RSY} where RSB occurs when
$O(u^2)$ terms are included in the equilibrium solutions.  (The approach of
Read, Sachdev, and Ye was slightly different from ours, since they expressed
all terms in the action in terms of the Hubbard-Stratonovich field $Q^{ab}$,
rather than working directly with the spin correlations, as we do here).

An important feature in the aging regime is the occurence of triangle relations
between correlations at different time scales.  These triangle relations have an 
ultrametric structure analagous to that found in RSB.  At sufficiently large 
times $t_1$ and $t_2$, it is possible to express the correlation at intermediate times, 
$t_2 < t < t_1$, in terms 
of a function $f$, which depends only on the correlations and has no explicit
time dependence. \cite{CugKurB}  Mathematically

\begin{equation}
C(t_1,t_2) = f[C(t_1,t),C(t,t_2)] \; ,
\end{equation}
and there is an inverse function $\bar{f}$

\begin{equation}
C(t_1,t) = \bar{f}[C(t_1,t_2),C(t_2,t)] \; .
\end{equation}
Cugliandolo and Kurchan \cite{CugKurB,CugKurC} give a complete account
of these triangle relations and the properties of the functions $f$
and $\bar{f}$.  The function $f$ can have fixed points $a$, such that
$a = f(a,a)$.  Each of these fixed points constitutes a correlation
scale.  We make use of this approach to calculate the FDT violation
factor $X$, in the long time dynamics of the quantum rotor model.

The saddle point equations are written down to $O(u^2)$ 
in Section \ref{sec:saddle}.  
To obtain the 
appropriate equations in the long time regime, the 
response must be split into a stationary
part and an aging piece, as in Equations (\ref{eq:csplit}) and (\ref{eq:rsplit}).  
At long times, the contributions to the dynamical equations from the 
bath kernels should be negligible, since the system 
has been in contact with the bath sufficiently long for one time quantitites to
have reached a limit.  What is studied here is how the two time quantities 
evolve after the interaction with the bath has ceased.  We assume WEB
and WLTM, and treat integrals in the manner explained by Cugliandolo and
Lozano. \cite{CugLoz}  Hence the 
equations we wish to solve are (for $t_1 \neq t_2$)

\begin{eqnarray}
0 & = & -\frac{1}{2g}\frac{\partial^2}{\partial t_1^2} R(t_1,t_2) + 
\lambda_1 R(t_1,t_2) \nonumber \\
& & + \lambda_2 R(t_1,t_2) C(t_1,t_2)^2 - \frac{1}{3}\lambda_2 \hbar^2
 R(t_1,t_2)^3 \nonumber \\
& & + 2J^2\int_0^\infty dt \, R(t_1,t)R(t,t_2) \nonumber \\
& & + \lambda_4 \int_0^\infty dt \, C(t_1,t)^2 R(t_1,t)R(t,t_2) 
\nonumber \\
& & - \frac{1}{3}\lambda_4 \hbar^2 \int_0^\infty dt \, R(t_1,t)^3 R(t,t_2),
\end{eqnarray}
and

\begin{eqnarray}
0 & = & -\frac{1}{2g}\frac{\partial^2}{\partial t_1^2} C(t_1,t_2) 
+ \lambda_1 C(t_1,t_2) + \kappa_2 R(t_1,t_2) \nonumber \\
& & + \frac{1}{3}\lambda_2 C(t_1,t_2)^3 
- \frac{1}{9}\lambda_2 C(t_1,t_2)R(t_1,t_2)^2  \nonumber \\
& & + \kappa_4 C(t_1,t_2)^2 R(t_1,t_2) - \frac{1}{3}\kappa_4 
\hbar^2 R(t_1,t_2)^3 \nonumber \\
& & + 2J^2 \int_0^\infty dt \, \left[C(t_1,t)R(t_2,t) + R(t_1,t)C(t,t_2)\right] \nonumber \\
& & + \frac{1}{3}\lambda_4 \int_0^\infty dt \, C(t_1,t)^3 R(t,t_2) \nonumber \\
& & + \lambda_4 \int_0^\infty dt \, C(t_1,t)^2 R(t_1,t) C(t,t_2) \nonumber \\
& & - \frac{1}{3}\lambda_4 \hbar^2 \int_0^\infty dt \, \left[ 3R(t_1,t)^2 R(t_2,t) C(t_1,t) 
\right. \nonumber \\ & & \left. \hspace*{1cm}
+ R(t_1,t)^3 C(t,t_2) \right]  \; , 
\end{eqnarray}
where we drop the aging subscript (since the only parts of the correlation and response we work 
with here are the aging pieces).  The values for the coefficients 
are displayed in Appendix \ref{coeff}. 

The equations derived here are very similar to those that have been
written down for the soft spin SK model. \cite{CugKurB,Dotsenko} If we
assume that at sufficiently long times only terms that are invariant
under time reparametrizations remain (more explanation is given in
Section \ref{RpG}
\cite{CugLoz,MCshort,Somp,CugKurB,Dotsenko,Ginzburg,Ioffe}), then the
dynamical equations take the form

\begin{eqnarray}
\label{eqn1}
0 & = & \lambda_1 R(t_1,t_2) + \lambda_2 C(t_1,t_2)^2 R(t_1,t_2)  \nonumber \\
& & + 2J^2\int_0^\infty dt \, R(t_1,t)R(t,t_2) \nonumber \\ & & 
+ \lambda_4 \int_0^\infty dt \, C(t_1,t)^2 R(t_1,t) R(t,t_2) \; , \\
\label{eqn2}
0 & = & \lambda_1 C(t_1,t_2) + \frac{1}{3}\lambda_2 C(t_1,t_2)^3 \nonumber \\ 
& & + 2J^2 \int_0^\infty dt \, \left[ C(t_1,t)R(t_2,t)  +
R(t_1,t)C(t,t_2) \right]  \nonumber \\
& & +\frac{1}{3}\lambda_4 \int_0^\infty dt \, C(t_1,t)^3 R(t_2,t) 
\nonumber \\ &  &
+ \lambda_4 \int_0^\infty dt \, C(t_1,t)^2 R(t_1,t) C(t,t_2).
\end{eqnarray} 
None of the terms with coefficients proportional to $\hbar^2$ enter into these 
asymptotic equations, hence the dependence on quantum effects is only through
the coefficients $\lambda_1$, $\lambda_2$, and $\lambda_4$.  There are also 
classical terms (in the sense that they do not have a coefficient 
proportional to $\hbar^2$) that 
are not reparametrization invariant.

The equations above have extra terms relative to the terms considered by 
Cugliandolo and Kurchan, \cite{CugKurB} which are
those with coeffcient $\lambda_4$ --  we can recover their results when these
terms are unimportant.  The reason that we have these terms is that we include all terms 
of $O(u^2)$ before looking at the long time limit.   Following Cugliandolo 
and Kurchan we can convert the 
equations into a manifestly time reparameterization invariant form by introducing functionals 
$F[C]$ and $H[C]$

\begin{eqnarray}
F[C] & = & - \int_C^q dC^\prime \, X[C^\prime] \; , \\
H[C] & = & - \int_C^q dC^\prime \, {C^\prime}^2 X[C^\prime] ,         
\end{eqnarray}
and postulating a modified version of the FDT to relate the correlation and response

\begin{eqnarray}
\label{eq:modFDT}
R(t_1,t_2) & = & \beta X[C(t_1,t_2)]\frac{\partial}{\partial t_2} C(t_1,t_2)
\theta(t_1 - t_2)
\; , \nonumber \\
& = & \beta \frac{\partial}{\partial t_2} F[C(t_1,t_2)]\theta(t_1 - t_2) \; .
\end{eqnarray}
The relation (\ref{eq:modFDT}) is also invariant under time
reparameterization.  We follow a similar reasoning to Ref. \onlinecite{CugKurB} to
show that within an ultrametric scale

\begin{eqnarray}
X[a_1^*] = \frac{\lambda_2 a_1^*}{\beta}\frac{(2J^2 + \lambda_4 q^2)^{\frac{1}{2}}}{(
2J^2 + \lambda_4 {a_1^*}^2)^{\frac{3}{2}}} \; .
\end{eqnarray}
We also find that within an ultrametic scale $X[C] = X[a_1^*]$, and that the 
scales have vanishing measure and there is thus a continous ultrametric 
solution at correlation scale $a_1^*$, 
provided the following condition is satisfied

\begin{eqnarray}
{\displaystyle
\frac{2x_1 F[a_1^*] - x_2 - \lambda_2 {a_1^*}^2}{\lambda_2 a_1^* +
\beta\lambda_4 F[a_1^*](2a_1^* X[a_1^*] - F[a_1^*])} > 0,}
\end{eqnarray}
where 

\begin{eqnarray}
x_1 & = & \beta(2J^2 + \lambda_4 {a_1^*}^2), \\
x_2 & = & \lambda_1 + 2\lambda_4 \beta \int_{a_1^*}^q dC^\prime \, C^\prime 
F[C^\prime], \\
F[a_1^*] & = & \frac{\lambda_2}{\beta\lambda_4} \left[ 1 - 
\sqrt{\frac{2J^2 + \lambda_4 q^2}{2J^2 + \lambda_4 {a_1^*}^2}}\,\, \right].
\end{eqnarray}
The results here reduce to those found earlier for the classical SK model 
\cite{CugKurB} in the limit that $\lambda_4 \to 0$.  Hence the solution for the
FDT violation factor $X[C]$ is 

\begin{eqnarray}
X[C] = \frac{\lambda_2 C}{\beta}\frac{(2J^2 + \lambda_4 q^2)^{\frac{1}{2}}}{(
2J^2 + \lambda_4 C^2)^{\frac{3}{2}}} \; .
\end{eqnarray}

Note that if we define $q$ to be the value of the correlation for which 
$X[C] = 1$, then this implies $q = \beta/\lambda_2 = \frac{2}{3}
\beta/(u^2\chi_\infty).$

\subsection{Reparametrization symmetry}
\label{RpG}
In deriving Equations (\ref{eqn1}) and (\ref{eqn2}) we assumed that the
equations at long times are reparametrization invariant.  As shown recently,
\cite{MCshort} this assumption is justified, since at long times the
dynamical equations governing the system {\it flow} to a ``fixed point'' of
the reparametrization group at which the equations are invariant under time
reparametrization transformations.  The time reparametrizations are
transformations of the form

\begin{equation}
t \to \tilde{t} = h(t),
\end{equation}
where $h(t)$ is a differentiable function with $dh/dt \ge 1$ (so as to 
stretch time).  Transforming time also leads to transformation of the 
correlation functions -- for a two-time correlation function $G(t_1,t_2)$,
we define a transformation $G \to \tilde G$ such that 

\begin{eqnarray}
\tilde{G}(t_1,t_2) = \left(\frac{\partial h}{\partial t_1}\right)^{
\Delta^G_1}\left(\frac{\partial h}{\partial t_2}\right)^{\Delta^G_2} 
G[h(t_1),h(t_2)],
\end{eqnarray}
where $\Delta^G_{1,2}$ are the scaling dimensions of the correlator
$G$ under the rescaling of the time co-ordinates $t_{1,2}$.  We term
$\Delta_1^G$ and $\Delta_2^G$ the advanced and retarded scaling
dimensions respectively.  The ``fixed point'' dynamical equations are
a set of equations which are invariant under an R$p$G transformation,
i.e. if they are satisfied by $G$, then they are also satisfied by
$\tilde{G}$.  The idea of irrelevancy under R$p$G transformations,
introduced in Ref. \onlinecite{MCshort}, where certain terms in the long
time limit scale to progressively negligible perturbations about an
R$p$G fixed point, is the means used to obtain the dynamical equations
at long times.

\section{Discussion}
\label{sec:discussion}

We have studied spin glasses of quantum rotors from a purely dynamical
perspective. In the dynamical approach, the phase transition between the
paramagnetic and glassy phase can be identified by looking at the critical
slowing down of the dynamics. Alternatively, we also identify the transition
at the breakpoint where it is no longer possible to satisfy the QFDT relation
between correlators and response functions in the solution of the equations
of motion. We find that, if one insists on a TTI solution, it is necessary to
include a singular ($\delta$ function) piece in the correlation to balance
the fluctuation dissipation relation. The amplitude of this singular term is
proportional to $q$, and this simple minded TTI solution is the analog
of the usual replica symmetric equilibrium solution. 

For the glassy phase, we study the non-equilibrium dynamics in the aging
regime, where TTI is broken. We find that the relation between correlation
and response for the quantum rotor model has the same character as in the
classical case -- specifically the OEFDR has the same form, but different
coefficients from the classical case. We show that all of the terms in the
dynamical equations that are only present in the quantum version of the model
(with an explicit $\hbar$ dependence) are not invariant under time
reparametrizations, but instead they are R$p$G irrelevant and do not
contribute in the long-time limit. The quantum terms, however, alter the
short-time dynamics and consequently renormalize the coefficients of the
classical (R$p$G invariant) terms in the long-time dynamical equations.

These findings help us understand rather more precisely how the behaviour of
a spin glass system is modified by the introduction of quantum dynamics.
Bhatt \cite{Bhatt} has argued that the scalings at the spin glass transition
should be the same as in classical models, due to the different timescales
involved -- $\beta\hbar$ is the timescale for the quantum case (this
corresponds to a frequency of $\omega_q = 8 \times 10^{11} \, T \, {\rm Hz}$,
where $T$ is the temperature in Kelvin), which at finite temperature will be
much shorter than other relevant time scales in the problem. Another way to
argue this is to look at the QFDT, which in the low frequency (long time)
limit corresponds to the classical FDT. However, the glassy systems we
consider here are never at equilibrium, and hence the long time limit in the
non-equilibrium system is less obvious.  One suggestion is that the QFDT
should be generalized in a manner similar to the generalization to an OEFDR
that differs from the classical FDT by introducing an effective temperature
$\beta X$ as the argument of the hyperbolic tangent.  \cite{CugLoz} The
picture we discuss here and in recent previous work \cite{MCshort} is that at
long times, due to reparametrization invariance of the dynamical equations,
the dynamics become classical in nature (as we would have naively expected
from equilibrium); however, the effective temperature is different to that
obtained in the corresponding classical model, due to quantum effects
renormalizing coefficients.

In essence, we find that in models where quantum dynamics is introduced in a
model with trivial spin commutation relations such as the rotors (in contrast
to $SU(2)$ spins, for example) quantum effects are only important in
basically two aspects of the problem. The first is an equilibrium issue: the
redefinition of the phase boundary due to the addition of quantum
fluctuations, which further disfavors an ordered phase. The second has a
non-equilibrium character: the renormalization of the classical terms in the
aging dynamical equations, which translates into a renormalized effective
temperature in the OEFDR. In this sense, this class of quantum spin glasses
only correspond to ``renormalized'' classical systems. The situation for
truly $SU(2)$ quantum spins, however, is not clear and deserves further
work.

In addition to the quantum-classical connection that we partly address
in this paper, there are also interesting issues connected to the time
reparametrization invariance of the long time dynamics. The time
reparametrization of the classical equations of motion appears to be a
generic feature of classical spin glass models. The invariance in the
SK model was noticed by several authors,
\cite{Somp,CugKurB,Dotsenko,Ginzburg,Ioffe,Horner} and we naturally
encounter it in the rotor model as well. The form of the OEDFR seems
to be a generic consequence of the retarded and advanced scaling
dimensions of $C,R$, \cite{MCshort} which in turn may be fixed by the
form of the FDT. In this sense, it would be interesting to discover
whether it is possible to construct different R$p$G fixed points, with
different scaling dimensions.


\section{ACKNOWLEDGEMENTS}
We acknowledge the hospitality of the Institute for Advanced Study at
Princeton (C.~C), where this work was started, of Boston University
(M.~P.~K), and of the University of Houston (J.~Y.). 
Support was provided by the NSF
Grant DMR-98-76208, and 
the Alfred P. Sloan Foundation (C.~C).


\begin{appendix}
\section{TTI solutions for the response}
One can attempt a solution for the response of the form $R(\omega) =
P(\omega) + iQ(\omega)$, It is then possible to solve for $P$ and $Q$ and
determine the quantum critical point, by writing $x(\omega) = a(\omega) +
ib(\omega)$.  The solution is

\begin{eqnarray}
R(\omega) & = & -\frac{1}{4J^2}(a+ib) \pm \frac{1}{4J^2}\sqrt{\left(a^2 - b^2
 - 8J^2\right) + 2iab},
\end{eqnarray}
which can be separated into real and imaginary parts:

\begin{eqnarray}
P(\omega) & = & -\frac{a}{4 J^2} \pm \frac{1}{4J^2} \left(\left(a^2 - b^2 
- 8J^2\right)^2 + 4 a^2 b^2 \right)^{\frac{1}{4}} \nonumber \\
& & \times \cos\left(
\tan^{-1}\left(\frac{2ab}{\sqrt{a^2 - b^2 - 8J^2 }}\right)\right), 
\\
Q(\omega) & = & -\frac{b}{4J^2} \pm \frac{1}{4J^2}  \left(\left(a^2 - 
b^2 - 8J^2 \right)^2 + 4 a^2 b^2 \right)^{\frac{1}{4}} \nonumber \\
& & \times  \sin \left(
\tan^{-1}\left(\frac{2ab}{\sqrt{a^2 - b^2 - 8J^2 }}\right)\right).
\end{eqnarray}
We obtain the critical point when the square roots in the arguments of the 
sine and cosine vanish.

\section{Coefficients for the aging equations}
\label{coeff}

The coefficients that appear for the dynamical equations in the aging 
regime are summarized below:
\begin{eqnarray}
\lambda_1 & = &  \frac{1}{2}m^2 + u(1+M)q
+ 2(1+M)u^2 B_\infty + 4 J^2 \chi_\infty \nonumber \\
& & + \frac{3}{2}(1+M)u^2 \gamma^{21}_\infty 
- \frac{1}{2}(1+M) u^2 \hbar^2 \gamma^{03}_\infty ,  \\
\lambda_2 & = & \frac{3}{2}(1+M)u^2 \chi_\infty ,  \\
\kappa_2 & = & 2J^2 \gamma^{10}_\infty ,  \\
\lambda_4 & = & \frac{3}{2}(1+M)u^2 ,  \\
\kappa_4 & = & \frac{3}{2}(1+M)u^2 \gamma^{10}_\infty ,
\end{eqnarray}
and

\begin{eqnarray}
\gamma_\infty^{mn} & = & \lim_{t_1 \to \infty} \int_0^{t_1} dt \, C_{ST}(t_1 - t)^m 
R_{ST}(t_1 - t)^n \; , \nonumber \\
B_\infty & = & \gamma^{11}_\infty + q\chi_\infty + 
\lim_{t_1 \to \infty}\int_0^{t_1} dt \, R_{AG}(t_1,t) C_{AG}(t_1,t) .
\end{eqnarray}

\end{appendix}

%
%
%
%

\end{document}